\documentclass[11pt]{article}

\date{}

\usepackage{url}
\usepackage{cite}
\usepackage{plain}

\usepackage{wrapfig}
\usepackage{graphics}
\usepackage[english]{babel}
\usepackage[leqno]{amsmath}
\usepackage{amssymb}
\usepackage{verbatim}
\usepackage[mathscr]{eucal}

\usepackage{amstext}
\usepackage{makeidx}
\usepackage{amsthm}

\usepackage{hyperref}
\hypersetup{colorlinks,%
citecolor=red,%
linkcolor=blue,%
urlcolor=black}

\makeindex

\newtheorem{theorem}{Theorem} 

\newtheorem{proposition}{Proprieta'}
\newtheorem{definition}{Definizione}
\newtheorem{notation}{Nota}
\newtheorem{ex}{Esercizio} 
\newtheorem{esempio}{Esempio}

\newcommand{\vs}{\vspace{3mm}}
 
\newcommand{\no}{\noindent} 

\newcommand{\beq}{\begin{equation}} 
\newcommand{\eeq}{\end{equation}}

\newcommand{\bex}{\begin{ex}} 
\newcommand{\eex}{\end{ex}} 

\newcommand{\bese}{\begin{esempio}} 
\newcommand{\eese}{\end{esempio}} 

\newcommand{\bpro}{\begin{proposition}} 
\newcommand{\epro}{\end{proposition}}

\newcommand{\bthe}{\begin{theorem}} 
\newcommand{\ethe}{\end{theorem}}

\newcommand{\bnote}{\begin{notation}} 
\newcommand{\enote}{\end{notation}}

\newcommand{\bdefi}{\begin{definition}} 
\newcommand{\edefi}{\end{definition}} 

\newcommand{\bc}{\begin{center}} 
\newcommand{\ec}{\end{center}}

\newcommand{\mail}[1]{\href{unina:#1}{\texttt{#1}}}

\usepackage{pdfsync}

\author{Monica De Angelis\thanks{Univ. of Naples  "Federico II", Scuola Politecnica e delle Scienze di Base. Dip. Mat. Appl. "R.Caccioppoli", \newline
 Complesso Universitario Monte S. Angelo
Via Cintia - 80126, Naples, Italy.
\newline\mail{modeange@unina.it}}
}

\title{Mathematical contributions to the dynamics of the Josephson junctions: state of the art and open problems}

\begin{document}
\maketitle

\begin{abstract}
 Mathematical models related to some Josephson junctions are pointed out and attention is drawn to the solutions of certain initial boundary problems and to some of their estimates. In addition, results of rigorous analysis of the behaviour of these solutions when $ t \rightarrow \infty $ and when the small parameter $ \varepsilon $ tends to zero  are cited. These analyses lead us to mention some of the open problems. 

\vs \no {\bf{Keywords}}:{\hspace{2mm} Third order parabolic operator,\hspace{2mm} Fundamental solution, \hspace{2mm}Superconductivity,\hspace{2mm} Josephson junction}

\vs \no \textbf{Mathematics Subject Classification (2000)}\hspace{1mm} 82D55,\hspace{1mm}74K30\hspace{1mm} 35K35 \hspace{1mm}35E05
\end{abstract}


\vspace{3mm}


\section{Introduction }

 Our purpose is to:

 i) furnish a short review of the mathematical contributions to the dynamics of the Josephson junctions,
 
 ii)introduce some possible open problems.
 
  From the mathematical point of view, many descriptions of superconductivity phenomena have been developed  and an  important contribution has been given by  Brian David  Josephson. He predicted in 1962  the  tunnelling of superconducting Cooper pairs  	through an 	insulating barrier to pass  from one superconductor to another (Josephson effect).  
  He also predicted the exact form of the current and voltage relations for the junction (Josephson junction) \cite{j1}.  
   (Experimental work proved that his theory was  right, and Josephson was awarded the 1973 Nobel Prize in Physics.)

 The  flux-dynamics of a Josephson junction, i.e,  two layers of superconductors separated by a very thin layer of insulating material, can be described   by means of  Sine Gordon equation (SGE):

\begin {equation}   \label {11}
   u_{xx}  - u_{tt} =\sin u 
\end{equation}

\noindent where $ x $ denotes the direction of propagation, $ t $ is  time and  the variable  $u= u(x,t)$ represents the difference between the phases of the wave functions of the two superconductors. 

However, in dealing  with real junctions it seems  necessary to take into account other effects  such as losses and  bias. Therefore, many authors    prefer to consider  the  so-called perturbed sine Gordon equation (PSGE):

  

\begin {equation}    \label {12} 
 \varepsilon u_{xxt}
+  u_{xx}  - u_{tt}  -  a u_t  = \sin u - \gamma. 
\end{equation}

In this case, terms $ \varepsilon u_{xxt}$ and $ 
  a u_t $ represent respectively the dissipative normal electron current flow along and across the junction,  (longitudinal and shunt losses) while $\gamma $ is the normalized current bias \cite {bp}. The value's range for $a$ and
 $\varepsilon$ depends on the real junction. Indeed, there are
 cases with $0<a,\varepsilon<1 $   
 and,
  when the shunt resistance of the junction is  low, the case $a $ large with respect
 to 1 arises  \cite{bp,tin,c}.

In some cases, extra terms must be considered. For  example in a   semiannular or  in a S-shaped  Josephson junction,  when  an applied magnetic
field $ b $ parallel to the plane of the dielectric barrier is considered, the dynamic equation   is: 

\begin {equation}    \label {13} 
 \varepsilon u_{xxt}
+  u_{xx}  - u_{tt}  -  a u_t  = \sin u - \gamma - b \cos ( kx) 
\end{equation}

\noindent where the last term evaluates  a transient force on the trapped fluxons  and locates these ones at the centre  of the junction. \cite{bp,sk,sk6}. Moreover, if an annular junction, also provided with   a microshort, is considered, the vortex dynamics in a static magnetic field is modelled  with the general perturbed sine-Gordon equation (see,f.i. \cite{sbs}):

\begin {equation}    \label {14} 
 \varepsilon u_{xxt}
+  u_{xx}  - u_{tt}  -  a u_t  = [1-\delta(x) \mu] \sin u - \gamma - b \cos ( kx)  
\end{equation}

\noindent where $ \mu $ is the current density associated with the microshort.

Nowadays, in addition to rectangular or annular junctions,  many other geometries for Josephson junctions have been  proposed. For instance,  window Josephson junctions  (WJJ) (\cite{bc} and reference therein) or  exponentially shaped Josephson junctions (ESJJ). \cite{bcs2,cmc02,mda13,32}.  This type of junction  is only a particular  case of a structure covering a region  

\begin{equation}
0\leq x\leq L  \qquad  g_2(x) \leq y \leq g_1(x).  
 \end{equation}
\noindent Denoting by  

\begin{equation}
0< w(x)= g_1(x)-g_2(x)\ll1,
\end{equation}

\noindent the evolution of the phase inside the  junction is given by:

\begin {equation}    \label {17} 
 \varepsilon u_{xxt}
+  u_{xx}  - u_{tt}  -  a u_t  = \sin u - \Gamma(x) -  \frac{\dot w(x)}{w(x)} (u_x+\varepsilon u_{xt})+\eta_y \frac{\dot w(x)}{w(x)} 
\end{equation}

\noindent where $ \Gamma (x) = \frac{\eta _x |_{g_2}- \eta_x |_{g_1}}{w(x)} $ and $ \eta _x \,\,\, \eta _y$  is the normalized magnetic field respectively in the $ x $ and  $ y  $ directions \cite{cmc02}.
When one assumes  $ g_1(x) =-g_2(x) =  w_o \, e^{-\lambda x} $,  where $ \lambda$ is a constant that, generally, is less than one,   an  ESJJ is obtained. Moreover,   assuming that there is  no bias current so that $ \Gamma (x) =0 $ and $ \eta_y =0, $  the equation achieved  is the following

\begin {equation}    \label {18} 
 \varepsilon u_{xxt}
+\, u_{xx}  -  u_{tt}  - \varepsilon\lambda u_{xt} -\lambda  u_x  -  a u_t \, \, = \, \sin u  .
\end{equation}

The current due to the tapering is represented by   terms  $ \,\,\lambda \,u_x $ and $\,\lambda \, \varepsilon \,u_{xt} \,$. In particular  $ \lambda u _{x} $  characterizes the  geometrical force driving the fluxons from the wide edge to the narrow edge.  These junctions assure many advantages compared to  rectangular ones,  such as    a voltage which is not  chaotic anymore, but rather periodic  excluding, in this way,   some among the possible causes of large spectral width. It is also proved that  the problem of trapped flux can be avoided (see f.i.\cite{cmc02}).

There exist numerous applications of  Josephson junctions  especially as superconducting quantum interference device (SQUID), which  consists of a loop of superconductor 
with one or more Josephson junctions.  These devices are one of the most important applications of superconductivity. They are basically extremely sensitive sensors of magnetic flux. This peculiarity allows to  diagnose heart and/or blood circuit problems using  magnetocardiograms and even  to evaluate  magnetic fields generated by electric currents in the brain using  magnetoencephalography -MEG- \cite {bp}. SQUIDs are  also used in nondestructive testing as a convenient alternative to ultra sound or x-ray methods (\cite{bp} and reference therein). In geophysics, instead, they are used as gradiometers \cite{tin} or as gravitational wave detectors (\cite{c} and reference therein).
 SQUIDs play an important role in the study of the  potential virtues of  superconducting digital electronics, too \cite{f}.
 
 \section{Mathematical models and equivalences}

All equations previously considered   have something in  common. More precisely, if one denotes by ${\cal L}\,  $  the following linear third order parabolic operator:

\begin{equation}  \label{21}
{\cal L}=\varepsilon \partial_{xxt}\, - \, \partial _{tt} \, +\, \partial_{xx}-\, \alpha \partial_t  
\end{equation}

\noindent (\ref{11})-(\ref{14}) and (\ref{18})  can be expressed    by means of  the  unique equations:  
 
\begin{equation}  \label{22}
{\cal L} u\,= f(x,t,u).
\end{equation}

\noindent According to the meaning of $ f, $  numerous other examples of dissipative phenomena can be considered. For example, equation (\ref{22}) arises in  the motion of viscoelastic fluids or solids \{see \cite{renno1,renno2,renno,dare,r,dr1} and references therein\}.  In particular, in \cite{renno1,renno2},  an isotropic, intrinsically homogeneous body with a linear viscoelastic behaviour of creep type  is considered, and    the following operator P

\begin{equation}
P u= c^2 u_xx-u_tt - \int_0^t g(t-\tau) u_{\tau\tau}(x,\tau)\, d\tau =-f
\end{equation} 

\noindent is  largely examined. Indeed,  by means of  a  rigorous analysis done in 
\cite{renno1,renno2,renno}, among other things,  the fundamental solution   for any arbitrary value of $ g $ have been  determined and asymptotic properties, maximal  principle and  singular perturbations results are achieved,too.

Moreover, an operator like (\ref{22}) appears in the   study 
of viscoelastic plates with memory, when the relaxation function is given by an exponential function \cite{rf} and it can also be employed  in the analysis of phase-change problems for an extended heat conduction model \cite{dade12,dade13}.

 In addition, equation (\ref{22}) arises also in  heat conduction at low temperature \cite{r,str} and  in the propagation of localized magnetohydrodinamic models in plasma physics \cite{sbes}. Still,  it  is possible to find others in  \cite{renno5,gb,rb,ccdt,jd}.

 Then,  an equivalence  between the third order equation (\ref{22}), typical of Josephson junctions,  and biological phenomena has been pointed out  in \cite{acscott}. Indeed, let us consider  the   FitzHugh-Nagumo system (FHN): \cite{i,m1}

\begin{equation}     \label{23}
  \left \{
   \begin{array}{lll}
    \displaystyle{\frac{\partial \,u }{\partial \,t }} =\,  \varepsilon \,\frac{\partial^2 \,u }{\partial \,x^2 }
     \,-\, v\,\,  -\,a\,u\, + u^2\, (\,a+1\,-u\,)   \qquad  \,(0<a<1)\,  \,  \\
\\
\displaystyle{\frac{\partial \,v }{\partial \,t } }\, = \, b\, u\,
- \beta\, v\,.
\\

   \end{array}
  \right.
 \end{equation}

\noindent where $\, u(x,t)\,$  represents  a membrane  potential  of  a  nerve  axon  at  distance  x and  time t, and  $\,v(x,t)\,$  is a  recovery variable that  models the  transmembrane current.

This reaction-diffusion  model characterizes the theory of the propagation of  nerve impulses, and the connection between a third order equation like (\ref{22}) and the (FHN) system  can be realized  changing the first one into the second one under continuous parameter variations \cite{acscott}.

An equation that is able to  model  all these physical problems   has been introduced in \cite{dri} and it is represented by   the following  parabolic integro-differential  equation:

\begin{equation}   \label{24}
 {\cal L}_R\,  u\equiv  u_t -  \varepsilon  u_{xx} + au +b \int^t_0  e^{- \beta (t-\tau)}\, u(x,\tau) \, d\tau \,=\, F(x,t,u) \,,
   \end{equation}

Indeed,   it  has been proved that (\ref{24})   characterizes  both reaction diffusion models like the FitzHugh-Nagumo system and  superconductive models \cite{dri,mda10,mmec,m13}.

In particular, perturbed sine-Gordon  equation  (\ref{12})  can be obtained by (\ref{24})  as soon as one assumes

\begin{equation}   
 a \,=\,  \alpha \, - \frac{1}{\varepsilon} \, \,\quad\,\, b = \,  - \, \frac{a}{\varepsilon}  \,\,\quad \displaystyle \, \beta \,= \frac{1}{\varepsilon}\,\,
  \end{equation}
  
  \noindent  and $ F  $ is such that

\begin{equation}   \label{26}
F(x,t,u)\,=\, -\, \int _0^t \, e^{\,-\,\frac{1}{\varepsilon}\,(t-\tau\,)}\,\,[\, \, sen \, u (x, \tau)\,- \gamma\, \,]\, \, d\tau. 
\end{equation}
 
Furthermore, the integro-differential  equation (\ref{24}) is able to  describe   the evolution inside an exponentially shaped Josephson junction, too. Indeed,  as it has already been underlined in \cite{32},  assuming

  \begin{equation} \label{27}
\beta \, = \frac{1}{\varepsilon}\qquad b= \, \beta^2 \, (1-\alpha \, \varepsilon \,) \qquad  a \, \beta  = \,\, \frac{\lambda^2}{4} \,-\,b \end{equation}
\[ 
   F\,= \, - \int_0^t\, \, e^{-\, \frac{1}{\varepsilon}(t-\tau) } f_1 (x,\tau, u) \,d \tau, \,\]

\noindent with 

\begin{equation}  \label{28}
 f_1 =\, e^{-\frac{\lambda} {2}\,x\,} [\, \sin \,( e^{\,x\,\lambda /2\,}\,   u ) \, - \gamma], 
  \end{equation}
  
\noindent from the integro differential equation (\ref{24}) it follows:

\begin{equation}  \label{29}
\varepsilon  u_{xxt}\, - \,  u_{tt} \, +\,  u_{xx}-\, (\alpha \,+ \varepsilon \, \frac{\lambda^2} {4})  u_t\,\,-\,  \frac{\lambda^2} {4}\, u = \, f_1   
\end{equation}

\noindent

\noindent Therefore, assuming   $ e^{\frac{\lambda} {2}\,x\,} \,  u =\bar u,\,   $    (\ref{29}) turns into equation  (\ref{18}). 

\noindent In  (\ref{24}) the  kernel  $ e^{- \beta (t-\tau)}\, u(x,\tau) $   can be modified   as physical situations demand and in this way  many other physical phenomena could be described.
 The particular choice made here  is  due to describe  the  superconductive  and biological models considered. 

\section{Mathematical results}

There exist many significant analytic results concerning the qualitative analysis of  equations related to Josephson junctions and  many initial-boundary problems have been
discussed in a lot of papers (see \cite{r,bsd,dmm,gh,ddf,df13} and references therein).

A first analysis, where the fundamental solution is determined, concerns operator $  {\cal L} $ in case $ \alpha=0 $ \cite{dare,dare1}.  Later, in \cite{ddr5,ddr6}, the fundamental solution of the whole operator  $  {\cal L}  $ of (\ref{21})  is explicitly determined and various properties are analyzed. Estimates and properties of continuous dependence for the solution of initial value problem are determined, too.

Moreover, in  \cite {mda01}, in order to  deduce an   exhaustive asymptotic analysis, the 
Green function of the linear operator $  {\cal L}  $ of (\ref{21}) 
has been determined by Fourier series and by means of its properties, an exponential decrease of  solution related to the Dirichlet problem  is deduced.
And still by means of Fourier series, existence and uniqueness for  Dirichlet, Neumann  and pseudoperiodic  initial-boundary  conditions are achieved,too \cite{df13,ddf}.

The  Dirichlet problem is still considered with respect to  equation (\ref{18}) and in \cite{mda13} the problem is reduced to an integral equation with kernel $ G $  endowed  with rapid convergence and   exponentially vanishing as $ t $ tends to infinity. 
\noindent Indeed, let

\begin{equation}                                             \label{31}
\gamma_n=\frac{n\pi}{l},\  \quad  b_n=(\gamma_n^2\,+ \lambda^2/4\,), \quad  g_n=\frac{1}{2} (\alpha\,+\,\varepsilon\,b_n\,),\  
\  \omega_n=\sqrt{g_n^2- \,b_n\,}
\end{equation}

\noindent and

\begin{equation}                                \label{32}
G_n(t)= \,\,\frac{1}{\omega_n}\,\,e^{-g_n\,t}\,\,
sinh(\omega_nt),
\end{equation}

\noindent  the Green function  is given by

\begin{equation}                                                \label{33}
G(x,t,\xi)=\frac{2}{l}\,\, e^{\frac{\lambda\,}{2\,}\,x}\,\,\sum_{n=1}^{\infty}\,
G_n(t) \  \  sin\gamma_n\xi\  \ sin\gamma_nx.
\end{equation}

The initial boundary problem with Dirichlet conditions is analyzed and an appropriate analysis implies results on the existence and uniqueness of the solution.

That is, indicating by 

\[
\,   \Omega_T \, \equiv \{\,(x,t) : \, 0\,\leq \,x \,\leq L \,\,;  \ 0 < t \leq T \,\}, \]

 \noindent the following  initial boundary problem   
 
  \begin{equation}          \label{34}
  \left \{
   \begin{array}{ll}
    & (\partial_{xx} \, - \,\lambda\,  \partial _x\,)\,\,(\varepsilon
u _{t}+ u) - \partial_t(u_{t}+\alpha\,u)\,=F(x,t,u),\ \  \
       (x,t)\in \Omega_T,\vspace{2mm}\\
   & u(x,0)=h_0(x), \  \    u_t(x,0)=h_1(x), \  \ x\in [0,L],\vspace{2mm}  \\
    & u(0,t)=g_1(t), \  \ u(l,t)=g_2(t), \  \ 0<t \leq T.
   \end{array}
  \right.
\end{equation}

\noindent   for $ g_1=g_2=0 $   admits the following integral equation:

\begin{equation}                                          \label{35}
 u(x,t)=
(\partial_t+\alpha+\,\varepsilon\,\lambda \partial_{x}\,-\varepsilon\partial_{xx})\int_{0}^{L} h_0(\xi) e^{-\frac{\lambda \xi}{2}}
G(x,\xi,t) d\xi 
\end{equation}

\[+ \int_{0}^{L} h_1(\xi) e^{-\frac{\lambda \xi}{2}} G(x,\xi,t) d\xi
+\int_0^ t d\tau\, \int_0^L G(x,\xi,t-\tau) 
e^{-\frac{\lambda \xi}{2}} F(\xi,\tau,u(\xi,\tau))d\xi,\]

So,  a priori estimates, continuous dependence and asymptotic behaviour of the solution, are deduced, too.

When boundary data are non null, equivalence between the integro-differential equation (2.4) and the equation describing the evolution inside an (ESJJ) has been considered in order to achieve explicit estimates of boundary contributions related to the Dirichlet problem.
Operator  $  {\cal L}_R $ of  (\ref{24})  has been extensively examined  and the fundamental solution $ K $  with  many of its properties have  been  determined in \cite{dri}. 

More in detail, if  $\, a,\,b,\, \varepsilon, \, \beta \, $ are positive constants, $\, r\,= |x| \, / \sqrt \varepsilon \, \, $ and $ J_n (z) \,$    denotes the Bessel function of first kind and order $\, n,\,$ let us  consider the function

  \begin{equation}     \label {36}
K(r,t)= \frac{e^{- \frac{r^2}{4 t}}}{2 \sqrt{\pi  \varepsilon t } } e^{-at}-  \frac{1}{2}  \sqrt{ \frac{b}{\pi \varepsilon }} \int^t_0  \frac{e^{- \frac{r^2}{4 y}- ay}}{\sqrt{t-y}}  e^{-\beta ( t -y)}  J_1 (2 ,\sqrt{by(t-y)\,})dy.
 \end{equation}
The following theorem has been proved:

\begin{theorem}

 The function  $\  K \, $  has the same  basic properties of the fundamental solution of the heat equation, that is:

 $ \,\,K(x,t) \, \, \in  C ^ {\infty} \,\,\,\,$ for \,$\,\,\, t>0, \,\,\,\, x \,\,\, \in \Re. $

  For fixed $\, t\,>\,0,\,\,\, K \,$ and its derivatives are exponentially vanishing as fast as $\, |x| \, $  tends to infinity.

 For any fixed $\, \delta \,>\, 0,\, $  uniformly for all $\, |x| \,\geq \, \delta,\, $ it results:

\begin{equation}                    \label{37}
 \lim _{t\, \downarrow 0}\,\,K(x,t)\,=\,0,
\end{equation}

 For $\,t\,>\,0,\,$ it is  $\,\,\, {\cal L}_R\,K    =\, 0.  \,\,$

Moreover, it results

\begin{equation}               \label{38}
|K(x,t)| \, \leq \, \frac{e^{- \frac{x^2}{4 \varepsilon \,t}\,}}{2\,\sqrt{\pi \varepsilon t}} \,\, [ \, e^{\,-\,at}\, +\, b t \,\,\,\frac{e^{\,-\,a t}\,-\,e^{\,-\beta\,t}}{\,\beta\,-\,a\,}\,\, ] 
\end{equation}
\hbox{}\hfill\rule{1.85mm}{2.82mm}
\end{theorem}

Previous estimates show, as well, that $\, K\,$  exponentially decays to zero as t increases.
These and other properties also allowed to prove in \cite{32}  numerous  properties of the following function which is similar to {\em theta functions}:

\begin{equation}     \label{39}
\theta (x,t)=  K(x,t) + \sum_{n=1}^\infty [\, K(x +2nL,t)  +  K ( x-2nL, t)]  =  
\end{equation}
\[ \sum_{n=-\infty }^\infty  K(x +2nL,t).\]
So that, denoting by 
  
\[G(x,\xi, t) \, = \,  \theta \,(\,|x-\xi|,\, t\,)\,- \,  \theta \,(\,x+\xi,\,t\,),
 \] 
 
\noindent  and

\[F(x,t,u) =  e^{-\frac{\lambda} {2}\,x\,} \,\,\biggl [ \int_0^t\, \, e^{-\, \frac{1}{\varepsilon}(t-\tau) } [\, \sin \,( e^{\,x\,\lambda /2\,}\,   u ) \, - \gamma] \,d \tau \,\,-\, \, h_1(x) \,\, e^{-\frac{t}{\varepsilon }}\biggr]\]

\noindent it has been proved that  problem  (\ref{34}) admits the following integro equation:

 \begin{equation}   \label{310}
 u( x,t) = \int^L_0  G(x,\xi, t)  e^{-\frac{\lambda} {2}x} h_0(\xi)d\xi  
+\int^t_0 d\tau\int^L_0  G(x,\xi, t)  F(\xi,\tau,u(x,\tau))d\xi \end{equation}
\[-2  \varepsilon \int^t_0 \theta_x (x,\, t-\tau)  g_1 (\tau )d\tau + 2 \varepsilon \int^t_0 \theta_x (x-L, t-\tau)  e^{-\frac{\lambda L}{2}} g_2 (\tau )d\tau.\]

  Besides,  a priori estimates  and asymptotic properties  have proved that when   $ t $ tends to infinity,  the effect due to the initial disturbances $\, (\,h_0, h_1\,) \, $ is  vanishing,  while  the  effect of the non linear source is bounded for all $ t. $ Furthermore, for large t, the  effects due to  boundary disturbances  $ g_1, g_2 $     are null or at least    everywhere bounded.
  
Indeed, if $ h_0 = h_1=0 $ and $ F=0, $ the following theorem holds:  

\begin{theorem}
 When   $ t $ tends to infinity and  data $ g_ i  \,\,\ (i=1,2) \,\,$  are two  continuous functions  convergent for large $ t$,  one has:  

\begin{equation}             \label{311}                                                 
 u  \, = g_{1,\infty}\,\,\frac{\sinh \sigma_0  \,\,(L-x)}{\sinh\  \sigma_0  \, L }  + g_{2,\infty}\,\,\frac{\sinh \sigma_0  \, x}{\sinh\  \sigma_0  \, L }
 \end{equation}
\noindent  where $ \sigma_0 =  \frac{\lambda}{2} $ and $ g_{i,\infty }= \lim_{t \to \infty } g_i , \,\, (i=1,2)$. Otherwise, when  $ \dot g_i  \in L_1[0,\infty] \,\,\, (i=1,2)$    too,
the effects determined by boundary disturbance  vanish. 
\hbox{}\hfill\rule{1.85mm}{2.82mm}
\end{theorem}

Another aspect  frequently highlighted in many papers is that
 the linear third order operator $  {\cal L}  $  is an example of wave operator perturbed by higher order viscous terms.  
The behaviour of solution of (\ref{22}) when $ \alpha =0 $,
  has been analyzed in various applications of artificial viscosity method \cite{nay,kl}.  Moreover, in  \cite{dmr},  when $ \varepsilon  $ is vanishing, the  interaction between diffusion effects and pure waves  has been evaluated  by means of slow time $ \varepsilon t  $ and fast times $ t/\varepsilon $.  These aspects are also  analyzed in \cite{dr1}  referring to the strip problem for equation (\ref{22}) with a  linear source term $ f $ or in \cite{renno,renno10}

Also equation (\ref{18}) can be considered as  a semilinear hyperbolic equation perturbed by viscous terms  described by  higher-order derivatives with small diffusion coefficients $ \varepsilon $. In  \cite{df213}, the influence of the dissipative terms has been estimated proving  that  they are both  bounded  when $ \varepsilon $ tends to zero and   when time tends to infinity, giving a mathematical proof of what has been observed in \cite{bcs2}.

As for  explicit solutions, an extensive literature exists, and more recently, various classes of solutions for (SGE) have been
determined.(see,f.i,  \cite{adc,xl}). 
Furthermore, when  $\varepsilon =0 $,   some  travelling-wave solutions  for (\ref{12}) have been  obtained both  for $|\gamma|$ not larger than 1 and  for $|\gamma|>1$ \cite{fiore}.
Still when  $ \varepsilon =0,$ some classes of explicit solutions have been determined for equation (\ref{18}),too \cite{df213}.

\section{Open problems}

In light of what has been stated  until now, many   open problems   can be highlighted. 

It  would be interesting, for example, to study equation (\ref{12})   when   interface conditions for the phase (and its normal gradient)   are added, connecting, in this way, with the problems of window Josephson junctions (WJJ)  when  the influence of an external magnetic field  must be considered \cite{bcf2}. Indeed, letting   $ \varepsilon =0 $, (\ref{12}) exactly  recalls one of the  equations  usually considered for  (WJJ). 

When, on the other hand,   $ \varepsilon  $ is not  vanishing, a viscous term,   represented   by the third order term, appears.  So that, it would  be  interesting   to give an estimate of the diffusive effects  due to  the $ \varepsilon $-term,too.

Moreover, according to the analogy between superconductor equations and  reaction-diffusion models, the Robin boundary  problem  would be  considered in order to achieve results  for  many biological phenomena, too \cite{rio,rio2}.

Besides,  as for  analysis on asymptotic effects due to  the boundary perturbations  related to equation (\ref{18}), as it has been pointed out,   the Dirichlet boundary  problem  has already been  considered in \cite{32}. So, the evaluation could be  extended to other boundary  problems, such as, for instance, Neumann  and  mixed ones. 

Of course, in order to  achieve  estimates for  other more  significant physical problems, this analysis and  many other estimates could be carried out for solution of equation  (\ref{13}) and for equations like (\ref{14}) where  the presence of a gap in the vacuum chamber is considered,too \cite{gh}   
    


The analysis conducted so far required that in (\ref{24}) constants $\, a,\,b,\, \varepsilon, \, \beta \, $ were all positive. This can be valid if we look for an analogy with an (ESJJ), but excludes application of (\ref{24}) to some other junctions. Therefore it would be interesting to extend the analysis of operator $  {\cal L}_R\, $ for any value of $\, a,\,b,\, \varepsilon, \, \beta. \, $

Finally a qualitative analysis of operators should be made in case $ \varepsilon, \alpha, \lambda  $  were not constant( see, f.i the case $a= 1- \cos u  $ \cite{bp}).

\section{Conclusion}
The state of the art  proves that many significant analytic results concerning the qualitative analysis of equations related to Josephson junctions have been obtained  and many initial-boundary problems have been discussed. However other many  important open problems may be considered and solved.

\bc{Acknowledgements}\ec
This paper has been written under the auspices of G.N.F.M. of INDAM.

\end{document}